\documentclass{aa}
\usepackage{graphicx}
\usepackage{txfonts}
\begin{document}
\authorrunning{G. Pugliese et al.}
\titlerunning{GRB030725 red optical afterglow}

\title{The red optical afterglow of GRB 030725}

\author{G. Pugliese \inst{1}, P. M{\o}ller \inst{1}, J. Gorosabel \inst{2,3}, 
B. L. Jensen \inst{4}, J. P. U. Fynbo \inst{4}, \\ J. Hjorth \inst{4}, 
S. F. J{\o}rgensen \inst{4}, B. Monard \inst{5}, \and C. Vinter \inst{4}}

\offprints{G. Pugliese, \email{gpuglies@eso.org}}

\institute{European Southern Observatory (ESO), Karl-Schwarzschild-Str. 2,
D-85748 Garching bei M\"unchen, Germany
\and
IAA-CSIC, PO Box 03004, 18080 Granada, Spain
\and 
Space Telescope Science Institute, 3700 San Martin Drive, Baltimore, MD 21218, USA
\and
Niels Bohr Institute, University of Copenhagen, Juliane Maries Vej 30, DK-2100 
K{\o}benhavn {\O}, Denmark
\and 
Bronberg Observatory, CBA Pretoria, PO Box 11426, Tiegerpoort 0056, Rep South Africa}

\date{Received ; accepted}

\abstract{We present a photometric study of the optical counterpart of the long-duration 
Gamma Ray Burst (GRB) 030725, which triggered the HETE FREGATE and WXM instruments on July 
25th, 2003, and lasted more than 160s. An optical counterpart was identified at the Bronberg 
Observatory in South Africa about 7 hours after the burst occurred. The optical afterglow 
(OA) was observed between 4 and 15 days after the burst with the 1.54m Danish telescope at 
La Silla in the {V}, ${\rm R_{c}}$, and ${\rm I_{c}}$ bands. We fit a broken power law to 
the data and determine a break time in the light curve between 16 hours and 4.7 days after 
the first detection of the burst. The decay slope is ${\alpha}_1 = -0.59^{+0.59}_{-0.44}$ 
before and $\alpha_2 = -1.43 \pm 0.06$ after the break. A bump may be present in the light 
curve, only significant at the 2${\rm \sigma}$ level, 13.9 days after the main burst. The 
spectral slope of the OA, measured 12 days after the burst, is $-2.9 \pm 0.6$ , i.e. it falls 
in the extreme red end of the distribution of previous OA spectral slopes. Observations of 
the field 8 months after the burst with the EMMI instrument on the NTT telescope (La Silla) 
resulted in an upper limit of ${\rm R_{c}}$=24.7 mag for the host galaxy of GRB 030725. The 
OA of GRB 030725 was discovered at a private, non-professional observatory and we point out 
that with the current suite of gamma ray satellites, an effort to organize future contributions 
of amateur observers may provide substantial help in GRB light curve follow up efforts.
\keywords{Gamma rays: bursts -- Stars: Supernovae: General -- Techniques: photometric}}

\maketitle
%

\section{Introduction}

Gamma Ray Bursts (GRBs) are powerful explosions in the ${\rm \gamma}$-ray band, with an 
emission which can peak up to a few MeV and a duration that varies between a few seconds up to 
hundreds of seconds (long GRBs). Since 1997, when the first optical counterpart was detected 
(\cite{Costa}; \cite{van1}), several satellites have been launched to study GRBs and for half 
of the well localized GRBs, an optical afterglow (OA) has been discovered.  As of today, there 
are about 60 detected GRB afterglows in the optical/IR bands. Detailed studies of GRB optical 
counterparts provide unique information on the characteristics of the medium in which the 
afterglow evolves, the jet structure of the emitting region and the energy associated with the 
main burst. Nevertheless, a larger and well defined statistical sample of well observed optical 
afterglows is necessary in order to address some still open questions concerning the universal 
structure of the jet-like emitting region and a quasi-standard energy reservoir (\cite{frail}; 
\cite{Bergerb}; \cite{Ghirlanda}), and the environment in which GRBs occur (\cite{yost}).

With the current suite of satellites, HETE-2, INTEGRAL and Swift, the rate of detected GRBs 
will reach an average of about 1 GRB every 3 days. Unless, however, the number of hours 
allocated to ground based optical follow-up will show a matching increase (about a factor of 
10), the build-up of the statistical sample will not proceed with the necessary speed, and 
many suitable afterglows will pass unobserved. Such a large increase is not realistic, and 
we must either explore alternative solutions or accept the loss of this opportunity to 
significantly increase our statistical sample. 

One possible alternative is to appeal to non-professional astronomers. The commonly used term 
``amateur astronomers'' does not properly describe the high quality of the equipment which 
is often available at non-professional observatories, and historically amateurs have made 
important contributions in the fields of Solar System and variable star studies. The large 
number of existent small private observatories could make significant contributions to GRB 
research in the fields of identification and high frequency early light curve sampling, the 
latter requiring that a suitable calibration of their data is available. In such a scheme 
a statistical sample of well documented GRB light curves could be created faster, and the 
professional GRB community would be free to be more selective in their follow-up strategy.
GRB 030725 is the first GRB optical afterglow to be discovered from a non-professional 
observatory, and therefore serves as evidence that there is a valuable potential in the
collaboration between amateurs and professionals in this field.

GRB 030725, a bright long hard burst, was detected by the FREGATE and WXM instruments on 
board of the HETE-2 satellite on July 25th, 2003 (\cite{Shiras}). It had a fluence of $\sim 
2 \times 10^{-5}$ ${\rm erg \, cm^{-2}}$ in the 30-400 keV band, and a fluence of $\approx$$2 
\times 10^{-5} \mathrm{erg \, cm}^{-2}$ in the 7-30 keV band. It showed a two peak shape, 
each peak with a FRED (Fast Rise Exponential Decay)-like structure. The first peak lasted 
about $\sim$40 s, and after about 160 s the second peak occurred with a duration of$\sim$10 
s. No detection of an afterglow in the X-ray or radio bands has been reported. 

In this article we present the results of the photometric ana\-ly\-sis of the optical 
afterglow of GRB 030725. The data were collected between 0.30 and 14.87 days after 
the main burst. We also observed the same field 8 months after the GRB occurred and 
obtained a limit on the magnitude of the GRB 030725 host galaxy in the ${\rm R_{c}}$-band.

\section{Observations and data reduction}

The optical afterglow of GRB 030725, located at R.A.= $20^{h}33^{m}59^{s}.47$ and Dec. = 
$-50^{o}40^{\prime}56^{\prime\prime}.26$ (J2000), with an error of $\pm0.1"$, was identified 
with the Bronberg 0.3m telescope approximately 7 hours after the trigger (Monard 2003), and a 
total of three exposures were obtained during the first night. Table 2 provides a log of those 
observations as well as of the later observations obtained with the 1.54m Danish telescope 
(1.54mD) on La Silla (\cite{Vinter}). Observations at the 1.54mD commenced 3.88 days after 
the burst during a period of adverse seeing conditions, and continued over the following 11 
nights.

The Bronberg telescope is equipped with a direct CCD camera (SBIG ST-7E) with 765 x 510 pixel 
CCD, providing an 18 arcmin E-W $\times$ 12.5 arcmin N-S field of view, with a pixel scale 
of 1.52 arcsec per pixel. The camera operates in unfiltered mode only. 

The DFOSC on the 1.54mD is equipped with the MAT/EEV CCD 44-82 "Ringo" camera, a pixel scale of 
0.39 arcsec/pixel, and a corresponding field size of $13.7' \times 13.7'$. Because the Bronberg 
data were unfiltered we needed to define an accurate calibration sequence with the 1.54mD in 
order to combine the two data sets. We therefore first describe the reduction and calibration 
of the 1.54mD data, but will subsequently return to the 0.3m data.

\begin{figure}[t]
\centering
\includegraphics[width=6.5cm,angle=270]{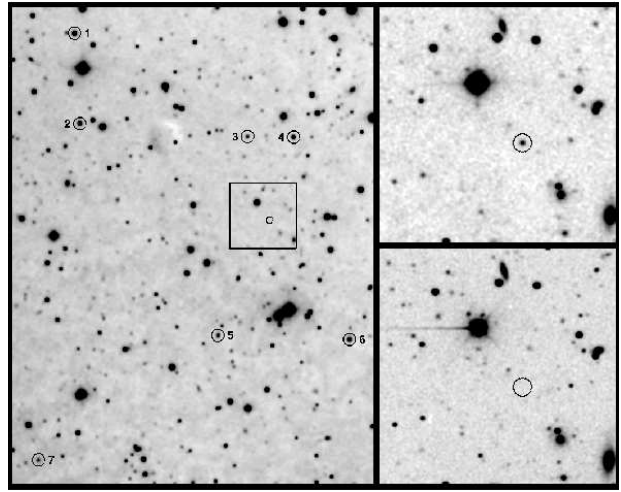}
\caption{{\bf Left:} The field of the GRB 030725 OA observed with the 1.54mD telescope 
          on La Silla. The numbered circles indicate the standard stars used for the 
          photometry. Their magnitudes are shown in Table 1.
         {\bf Upper right:} Zoom of the field (marked box in the left image) around the GRB 
          030725 OA observed with the 1.54mD telescope on La Silla 3 days after the 
          burst occurred. OA marked with a circle. 
         {\bf Lower right:} The 3.5m ESO-NTT image of the field of the afterglow obtained 8 
          months after the trigger. Neither OA nor host galaxy is visible, its position is 
          indicated with a circle.}
\label{GRB1}
\end{figure}

The attempt to detect the host galaxy of GRB 030725 was performed with the NTT telescope at 
La Silla. Fig.~\ref{GRB1} shows (left panel and right upper panel), the field of the GRB 
030725 observed with the 1.54mD telescope on La Silla 3 days after the burst occurred. 
The right lower panel shows the same field observed with the 3.5m ESO-NTT telescope on La 
Silla, 8 months after the trigger. 

\begin{table*}
\caption[b]{Magnitudes of the reference (ref) stars used for the photometry in the ${\rm R_{c}}$ 
band.}
\label{cal}
$$
\begin{array}{cccccccc}
\hline
\noalign{\smallskip}
 & {\rm ref1} & {\rm ref2} & {\rm ref}3 & {\rm ref4} & {\rm ref5} & {\rm ref6} & {\rm ref7}  \\
\noalign{\smallskip}
\hline\hline
\noalign{\smallskip}
 {\rm Mag} & 16.12 & 16.31 & 17.97 & 16.78 & 17.96 & 16.20 & 19.08  \\
 {\rm Mag. Error} & 0.04 & 0.03 & 0.04 & 0.04 & 0.03 & 0.04 & 0.03  \\
\noalign{\smallskip}
\hline
\end{array}
$$
\end{table*}

\subsection{1.54mD data reduction and calibration}

The data from the 1.54mD telescope in the {V}, ${\rm R_{c}}$, and ${\rm I_{c}}$ bands 
were processed using standard IRAF\footnote{IRAF is distributed by the National Optical 
Astronomical Observatory, which is operated by the Association of Universities for 
Research in Astronomy, Inc., under cooperative agreement with the National Science 
Foundation.} tasks for bias-subtraction and flat-fielding. The photometry was performed 
using the DAOPHOT package in IRAF (\cite{Stetson1}; \cite{Stetson2}). 

We performed aperture and PSF photometry on both the optical afterglow (OA) and 7 isolated, 
bright unsaturated field stars which were used as reference stars for the relative photometry. 
These stars are shown in the left panel of Fig.~\ref{GRB1} and their corresponding magnitudes 
are shown in Table 1. To calibrate the light curve of the OA in the ${\rm R_{c}}$-band, two 
standard fields, MarkA and SA110, were chosen from the Landolt catalogue of standard stars. 
These two fields were observed in the {B}, {V}, ${\rm R_{c}}$ and ${\rm I_{c}}$ bands. The 
corresponding calibrated magnitudes of the optical afterglow in the {V}, ${\rm R_{c}}$ and 
${\rm I_{c}}$ bands with the 1.54mD telescope are shown in Table 2. 

\begin{table*}
\caption[b]{Observing log (left) and photometric results (right) of the GRB030725 optical 
afterglow. ``Unf'' refers to the unfiltered data taken with the 0.3m telescope, while 
${\rm t - t_0}$ refers to the observing epoch. The first 5 data points in the table were 
collected in July 2003, while all the following data were collected in August 2003.}
\label{phot}
$$
\begin{array}{cccc||cccc}
\hline
\noalign{\smallskip}
{\rm Date} & {\rm Exp Time} & {\rm Seeing} & {\rm Telescope}  &  {\rm {t-t_0}} & 
 {\rm Filter} & {\rm Magnitude} & {\rm Mag. \, \rm Error} \\ 
{\rm (UT time)} & {\rm (sec)} & {\rm (arcsec)} & & {\rm (days)} & & & \\
\noalign{\smallskip}
\hline\hline
\noalign{\smallskip}
 25.7944 & 360   & 3.1 & {\rm 0.3m}  & 0.30  & {\rm Unf} & 18.59^{\mathrm 1 } & 0.21 \\
 25.8393 & 360   & 2.5 & {\rm 0.3m}  & 0.38  & {\rm Unf} & 18.67^{\mathrm 1 } & 0.21 \\
 25.9437 & 360   & 3.0 & {\rm 0.3m}  & 0.48  & {\rm Unf} & 18.90^{\mathrm 1 } & 0.21 \\
\cline{1-8}
 29.3355 & 5500  & 3.9 & {\rm 1.54mD} & 3.88  & {\rm R_{c}} & 21.20 & 0.04 \\
 30.3151 & 1800  & 4.4 & {\rm 1.54mD} & 4.83  & {\rm R_{c}} & 21.51 & 0.05 \\
 01.4347 & 1200  & 4.9 & {\rm 1.54mD} & 6.93  & {\rm R_{c}} & 22.46 & 0.55 \\
 02.2920 & 7800  & 4.9 & {\rm 1.54mD} & 7.85  & {\rm R_{c}} & 22.25 & 0.06 \\
 03.3268 & 7800  & 4.8 & {\rm 1.54mD} & 8.87  & {\rm R_{c}} & 22.45 & 0.08 \\
 04.7899 & 13800 & 5.7 & {\rm 1.54mD} & 10.34 & {\rm R_{c}} & 22.57 & 0.10 \\
 06.3535 & 3600  & 5.4 & {\rm 1.54mD} & 11.88 & {\rm R_{c}} & 22.96 & 0.10 \\
 07.3176 & 4200  & 4.6 & {\rm 1.54mD} & 12.87 & {\rm R_{c}} & 23.04 & 0.09 \\
 08.3116 & 7800  & 5.0 & {\rm 1.54mD} & 13.86 & {\rm R_{c}} & 22.87 & 0.17 \\
 09.3360 & 6600  & 4.2 & {\rm 1.54mD} & 14.87 & {\rm R_{c}} & 22.54 & 0.20 \\
 06.3211 & 3600  & 4.1 & {\rm 1.54mD} & 11.83 & {\rm   V  } & 23.61 & 0.15 \\
 07.3383 & 4800  & 4.5 & {\rm 1.54mD} & 12.85 & {\rm I_{c}} & 22.12 & 0.17 \\
 09.2839 & 3600  & 3.2 & {\rm 1.54mD} & 14.79 & {\rm I_{c}} & 22.38 & 0.19 \\
\noalign{\smallskip}
\hline
\end{array}
$$
\begin{list}{}{}
\item[]
$^{\mathrm{1}}$ ~{Unfiltered data calibrated to the ${\rm R_{c}}$-band}
\end{list}
\end{table*}

\subsection{The Bronberg data}

The first detection of the OT of GRB 030725 was obtained seven hours after the main burst 
occurred using unfiltered CCD imaging with an ``amateur'' telescope at the Bronberg 
Observatory. No photometric standards were obtained with the same instrument. Potentially 
the very wide response of unfiltered data could introduce severe colour terms which would 
make it difficult to combine the unfiltered data with the later $R_{c}$-band data. However, 
the large gap in time between the Bronberg and 1.54mD data renders the first three data 
points important in the determination of the decay parameters, and for this reason we went 
through a detailed calibration procedure aimed at determination of zero-point and colour 
terms of the unfiltered data. 

We followed a procedure very similar to that used for the photometric analysis of GRB 020813 
by \cite{li}. We selected 16 isolated stars which were unsaturated in both the 3 unfiltered 
images and the 4 short exposure data in the {B}, {V}, ${\rm I_{c}}$ and ${\rm R_{c}}$ bands. 
For each of these 16 stars the calibrated colours {B}-{V}, {V}-${\rm R_{c}}$, and 
${\rm R_{c}}$-${\rm I_{c}}$, as well as the unfiltered colours {B}-{Unf}, ${\rm R_{c}}$-{Unf}, 
${\rm I_{c}}$-{Unf}, and {V}-{Unf} were computed. It turned out that the smallest gradients 
were found for ${\rm R_{c}}$-{{Unf}}, indicating that the effective central wavelength of {Unf} 
is close to that of ${\rm R_{c}}$. Because the unfiltered colour ${\rm R_{c}}$ - {Unf} showed 
to be rather independent of each of the calibrated colours for all the selected stars, we 
concluded that the ${\rm R_{c}}$-band was the one which best reproduced the unfiltered data. 

The OA ${\rm R_{c}}$-band magnitudes inferred from the calibrated unfiltered data are shown in 
the upper part of Table 2.

\subsection{The host galaxy}

The study of GRB host galaxies is an important step to determine GRB redshifts and to study 
the environment in which GRBs occur (\cite{Jakob}). 

An attempt to detect the host galaxy of GRB 030725 was done on March 22, 2003, about 8 months 
after the main burst. The field in which the GRB OA was first detected was observed with the 
ESO NTT 3.5m telescope on La Silla. We used the red arm of the EMMI instrument in the 
${\rm R_{c}}$-band and obtained 3 exposures, each of 600 seconds, with a seeing of 0.6 arcsec. 
The data were processed using standard bias-subtraction and flat-field techniques and the 
photometry was computed using the same procedure described in the previous paragraph. There 
is no detection of the host galaxy and we compute a 2-sigma upper limit of 24.7 magnitudes 
in the ${\rm R_{c}}$-band for the GRB host galaxy.

\section{Light curve and spectral slope}

Many theoretical models have been developed to describe the spectral and temporal evolution 
of GRB afterglows (\cite{Sari}; \cite{Vanna}; see also \cite{Zhang} for a comprehensive review). 
Each of these models assumes that the afterglow is created when relativistic particles are 
accelerated at the shock front between the flow generated by the GRB explosion and the 
interstellar medium (ISM), and emit synchrotron radiation visible in different bands. The 
distribution of the relativistic electrons is assumed to be a power-law, $N(\gamma_{e}) \, 
d \gamma_{e} \propto \gamma_{e}^{-p} \, d \gamma_{e}$, where ${\rm \gamma_{e}}$ is the Lorentz 
factor. One of the main parameters which varies among all the theoretical models is the 
value assumed for $p$, the exponent of the electron distribution, whose range is generally 
chosen between 1 and 2.5, (\cite{van2}). Correspondingly, each model will obtain different 
power-laws for the decay of the afterglow, which is well represented by the formula $F_{\nu} 
\propto t^{\alpha} \nu^{\beta}$, with ${\rm \alpha , \beta} \ < 0$.

\subsection{The light curve: Comparison with theoretical models}

\begin{figure}
\centering
\hspace*{-1.0cm}
\includegraphics[width=10.3cm]{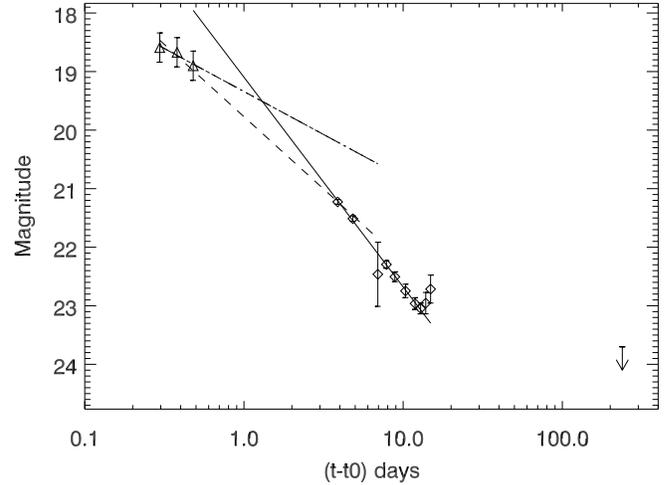}
\caption{Light curve of GRB 030725 in the ${\rm R_{c}}$-band. The solid line represents the fit 
        of the data from the 1.54 Danish telescope (diamonds) with a power law, assuming a break 
        in the light curve $3.88^{+ 1.11}_{- 0.48}$ days after the burst and no SN contribution. 
        The corresponding slope is equal to ${\alpha}_2 \simeq -1.43 \pm 0.06$. The dashed line 
        represents the fit of the first four data points with a power law and a corresponding 
        slope equal to ${\alpha}_1 \simeq -0.99 \pm 0.04$. The dotted-dashed line represents 
        the fit of the first 3 data points with a power law fitting with a slope of ${\alpha}_1 
        \simeq -0.59 \pm 0.67$ and an earlier break in the light curve, $1.30^{+ 2.32}_{- 0.14}$ 
        days after the burst. The arrow represents the upper limit to the magnitude of the 
        host galaxy from the NTT data.}
\label{GRB4}
\end{figure}

The light curve of GRB 030725 in the ${\rm R_{c}}$-band is shown in Fig.~\ref{GRB4}. The data 
points fall in two groups, where the second group (1.54mD data) covers the longest span in 
time, and has the smallest error bars. Except for a possible upturn, or ``bump'' (see below), 
seen in the last few data points, this part of the curve is well fitted by a power-law with 
slope ${\alpha}_2 \simeq -1.43 \pm 0.06$. The slope of the three early data points alone is 
less well determined (${\alpha}_1 = -0.59 \pm 0.67$) but this can be further constrained by 
considering the break time. 

It is clear that a break must have occurred, but we are left with two possibilities. In case 
the break occurred earlier than 3.88 days after the burst then the early power-law is the 
one stated above, with the additional limit that ${\alpha}_1 > -0.98$. If the break occurred 
later than 3.88 then the first 1.54mD data point falls within the pre-break light curve, and 
we find ${\alpha}_1 \simeq -0.99 \pm 0.04$. Both scenarios are consistent with the data and 
both fits are shown in Fig.~3. We conclude that the break occurred before 4.7 days, and that 
the early power law had a slope flatter than $-1.03$, i.e. $\rm{{\alpha}_1} = 
-0.59^{+0.59}_{-0.44}$. For the purpose of determining the break time we only consider 
negative values of $\rm{{\alpha}_1}$.

Breaks have been observed in the light curve of many afterglows (see \cite{Piran} for an 
extensive review), and one of the possible explanations is related to an emitting region with 
a jet-like structure. According to this model, as long as $1 / \Gamma < \theta_{jet}$, where 
$1 / \Gamma$ and $\theta_{jet}$ are the relativistic beaming factor and the opening angle of 
the emitting region respectively, it will not be possible to distinguish between a spherical or 
a collimated emitting region. But when $1 / \Gamma = \theta_{jet}$, then the jet structure will 
become visible and this will correspond to a break in the power-law decay of the light curve. 

In the case of GRB 030725, the lack of data between 0.48 and 3.88 days, did not allow an 
accurate estimate of the time of the break and the slope of the early power-law decay. Many 
models predict a change of slope between a few hours up to a few days after the main burst,
consistent with our upper limit of 4.7 days (observers frame). The allowed range of 
${\rm \alpha_1}$ reported above is in good agreement with slopes observed in other afterglows, 
$\rm{\alpha_1} \simeq -1$, while the slope of the light curve after the break is in the extreme 
flat end of the distribution of other afterglows $-2.5 \le \rm{\alpha_2} \le -1.5$, 
(\cite{Ander}). 

If the break in the light curve of GRB 030725 is related to the presence of a collimated 
outflow, then the jet structure appeared between 16 hours and 4.7 days after the main burst, 
and ${\rm p}$, the exponent of the power law distribution of electrons, should be less than 2. 
The lack of data during the first few days after the burst did not allow us to give a more 
accurate interpretation of this break.

\subsection{The light curve: A late time bump?}

Inspection of Fig.~\ref{GRB4} shows suggestive evidence of an ``upturn'' in the light curve 
after 13 days (observer frame). Such late time bumps in the light curves have been observed 
on several occasions, e.g. GRB 980326 which showed a red bump about 3 weeks after the burst 
(\cite{Blooma}); (\cite{cast}), GRB 011121 with a bump whose peak occurred between 13 and 
75 days after the first ${\rm \gamma}$-ray observation (\cite{Bloomb}), GRB 020405 with 
evidence of a red late bump between 10 and 25 days after the main burst (\cite{Price}), GRB 
021211 which showed the bump 27 days after the main burst (\cite{Valle}), and GRB 041006 
with evidence of a bump 4 days after the burst (\cite{Stanek}; see \cite{Zeh} for a 
comprehensive review). For these GRBs the bump features were identified as SN contributions. 

To test if our data support the presence of a SN we performed minimum ${\rm \chi^2}$ two 
component model fits to the light curve where one component was a power law (the GRB) and 
the other was one of three historical SN light curves redshifted to a set of different 
redshifts (SN1994I, SN1998bw, and SN2002ap (\cite{sode})). None of the models including
SN templates provided a significantly better fit than the fit of a pure power law, so we 
conclude that there is no significant evidence that the marginal (2 ${\rm \sigma}$) upturn 
after 13 days can be identified with a SN bump.


\subsection{Spectral slope}

The OT was detected in three bands ({V}, ${\rm R_{c}}$, ${\rm I_{c}}$) 12--13 days after the 
burst. On the assumption that the ${\rm \alpha_2}$ decay was monochromatic, we computed 
(including corrections for Galactic reddening), {V}-${\rm R_{c}} = 0.70^{+0.19}_{-0.17}$ and 
${\rm R_{c} - I_{c}} = 0.87^{+0.19}_{-0.21}$ (two sided errors computed using ML techniques, 
see Appendix A in \cite{fynbo1}). In order to compute the spectral slope ${\rm \beta}$ ($F_{\nu} 
\propto \nu^{\beta}$, ${\rm \beta} < 0$) we transformed the {V}, ${\rm R_{c}}$, ${\rm I_{c}}$ 
magnitudes to the AB system using the offsets to the AB system given in \cite{fuku} and fit a 
linear relation to corresponding values of ${\rm \nu_c}$ and AB. Here ${\rm \nu_c}$ is the 
central frequency for the filters. In this way we infer a spectral slope of ${\rm \beta} 
= -2.9\pm0.6$. Even allowing for the rather large error bar, this value is in the extreme 
red end of the distribution of spectral slopes, where the typical value is around 1 and a 
few limit cases around 2 (\cite{Simon}). We note that the OT colour was obtained 12 days 
post-burst, so we can interpret it as either the result of significant extinction due to 
dust in the host galaxy, or the result of a contribution from an underlying SN, or the GRB 
occurred at a high redshift.


\section{Summary and discussion}

Since the discovery of the first GRB optical counterpart in 1997, about 300 GRBs have been 
detected by several satellites, but only for about 25\% of them has an optical afterglow
been detected (from webpage: http://www.mpe.mpg.de/$\sim$jcg/grbgen.html). For even 
smaller percentage, about 15\% of the total detected GRBs, the redshift has been determined 
and only a few of these showed a late time bump in the optical light curve. 

GRB 030725 was a long hard GRB with a slope ${\alpha}_1 = -0.59^{+0.59}_{-0.44}$ of the early 
time optical light curve, a break between 16 hours and 4.7 days after the main burst, and a 
slope $\alpha_2 = -1.43 \pm 0.06$ after the break. On the assumption that the break is related 
to a collimated outflow, and using the relation between the opening jet angle and the time in 
which the break in the light curve occurs (\cite{Vanna}), we find jet opening angles of 
$7^{\circ}$ and $12^{\circ}$, at 16 hours and 4.7 days respectively. 

We detected a weak feature of a late time upturn in the light curve 14 days after the burst 
and tried to model it with the contribution from SNs at different redshifts, using SN1994I, 
SN1998bw, and SN2002ap as templates. We found that none of these SNs provided a better fit 
to the data than a simple power law. 

We did an attempt to detect the host galaxy of GRB 030725 about 8 months after burst, and 
we derived a 2-sigma upper limit of ${\rm R_{c}}$=24.7 mag for the GRB host galaxy. 

We determined the colours of GRB 030725 (${\rm V} - {\rm R_{c}} = 0.70^{+0.19}_{-0.17}$ and 
${\rm R_{c}-I_{c}} = 0.87^{+0.19}_{-0.21}$) corresponding to a spectral slope of ${\rm \beta} 
= -2.9\pm0.6$. This is significantly redder than the majority of GRB OAs (\cite{Simon}). 
The characteristics of the environment in which GRB occur are still an open question, 
(\cite{Bergera}; \cite{fynbo2}; \cite{ceron}; \cite{Tanvir}), but so far no GRB OA was found 
with such a steep spectral slope. This redder slope could be due to either high redshift, or 
contribution from an underlying SN, or dust absorption in the host galaxy. Some authors 
(\cite{Hjorth}; \cite{Prochaska}; \cite{Vreeswijk}) have shown that some GRB host galaxies 
have a rather low amount of dust and a rather low metallicity. The environment of GRB 030725 
could be very dense along the line of sight to the GRB itself. Further observations in IR to 
mm wavelength bands would be interesting as they would allow us to better understand the 
nature of this spectral slope. 

\begin{acknowledgements}
We thank IJAF for the allocation of observing time at the Danish 1.54 m telescope on La Silla. 
We also thank S. S. Larsen for the NTT data of the observation of GRB 030725 field in March 2004. 
G.P. thanks ESO DGDF fundings for supporting this research. The research of J. Gorosabel is 
supported by the Spanish Ministry of Science and Education through programs ESP2002-04124-C03-01 
and AYA2004-01515 (including FEDER funds). BLJ acknowledges support from the Instrument Centre 
for Danish Astrophysics (IDA) and the Nordic Optical Telescope (NOT).
\end{acknowledgements}


\end{document}